\documentclass[12pt,a4paper]{article}
\usepackage{amsmath}
\usepackage[dvips]{graphicx}
\usepackage{times}
\begin{document}
\begin{titlepage}
\title{On the possible decoherence observed at $\sqrt{s}=7$ TeV}
\author{S.M. Troshin, N.E. Tyurin\\[1ex]
\small  \it Institute for High Energy Physics,\\
\small  \it Protvino, Moscow Region, 142281, Russia}
\normalsize
\date{}
\maketitle

\begin{abstract}
We  briefly discuss  the possible decoherence observed at the LHC emphasizing that
the value of the average transverse momentum at  $\sqrt{s}=7$~TeV can be interpreted as a manifestation of the deviation
from the mechanism of the transient state of rotating matter. 
\end{abstract}
\end{titlepage}
\setcounter{page}{2}

It is well known  that the deconfined state, which has been discovered at RHIC,
reveals the properties of the perfect liquid, i.e. it is a strongly
interacting collective state \cite{rhic,rhic1,rhic2}.  Natural question arizes then: will the transient state retain its collective
nature at the LHC energies or not?

In the papers \cite{pt,v2,rv} the mechanism of hadron interactions based on collective properties of the transient matter
has been proposed. This mechanism uses 
the  geometrical picture of hadron collision at non-zero impact parameters
which implies that the generated massive
virtual  quarks in overlap region  obtain  large  orbital angular momentum
at high energies. Due to strong interaction
between quarks mediated by pions this  results in  the coherent rotation
of the quark system located in the overlap region as a whole  in the
$xz$-plane.  
This mechansim allows to explain, at least qualitatively, several experimental findings at the LHC energies
and among them an appearance of the ridge and anisotropic flows in $pp$--collisions \cite{v2,rv}.
It also allowed to get quantitave description of the energy dependence of the average transverse momentum \cite{pt}.

This point needs a few comments since it will be discussed further. First of all, the quantitave description mentioned above
includes experimental data for $pp$ and $\bar p p$ interactions obtained in the experiments at CERN (ISR, UA1), FNAL (E735, CDF)
and data for the energies $\sqrt{s}=1.8$ TeV and $\sqrt{s}=2.36$ TeV obtained at CERN LHC. We does not differentiate experimental
points relevant for the $pp$ and $\bar p p$ interactions and consider those processes to be identical in this respects. It has
firm experimental justification \cite{ul}. 
\begin{figure}[h]
\begin{center}
\resizebox{6cm}{!}{\includegraphics*{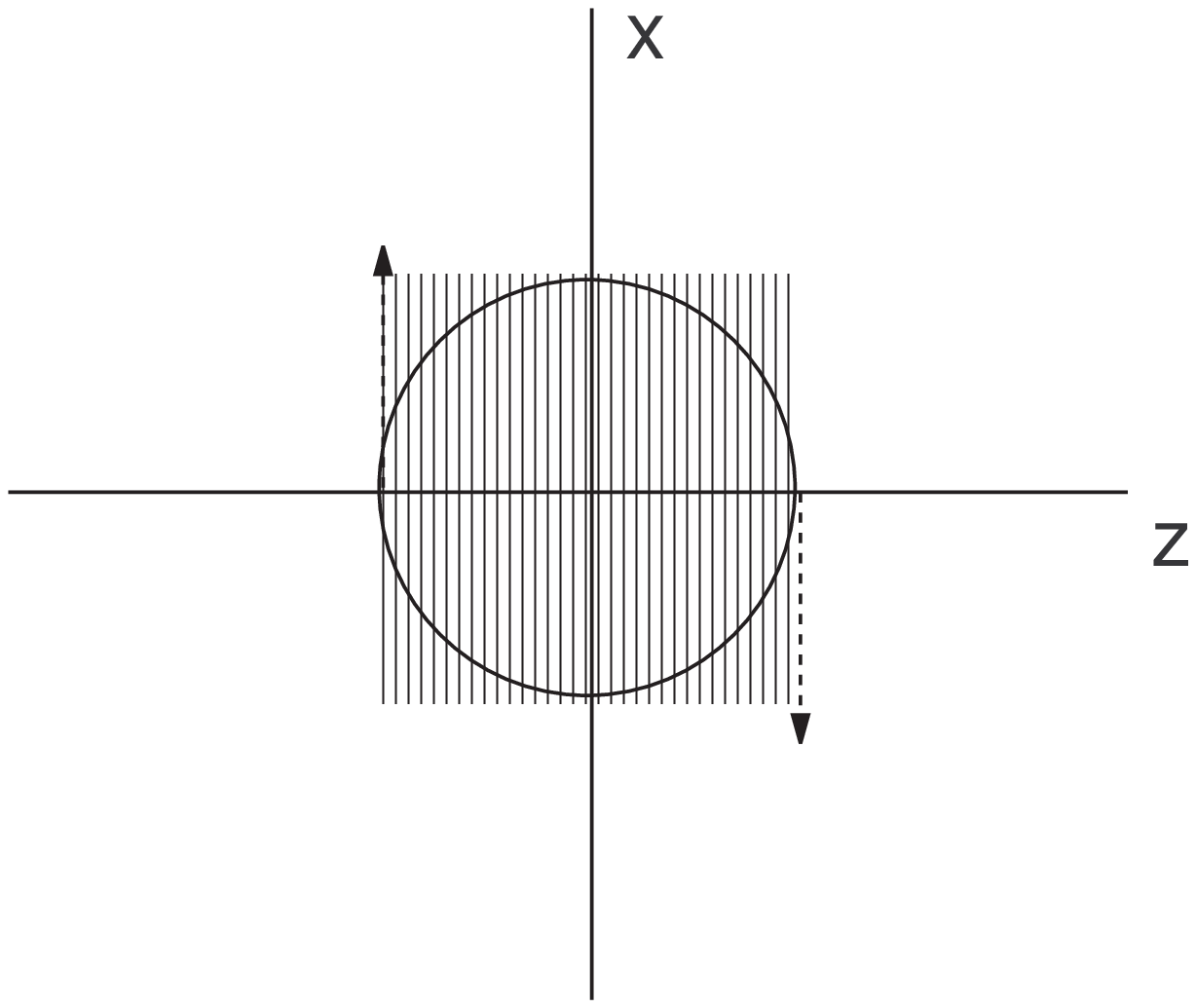}}
\end{center}
\end{figure}
Keeping this point in mind, a few notes on the assumed mechanism of transverse 
momentum generation should be recalled.
One should take into account that the rotation of the transient matter in the overlapping region 
 gives a contribution to the $x$-component of the transverse 
momentum (it does not contribute to the $y$-component of the transverse momentum), i.e.
\begin{equation} \label{ptx}
\Delta p_x=\kappa L(s,b), \; \Delta p_y=0,
\end{equation}
where $L(s,b)$ is the orbital angular momentum in the proton (or antiproton) collison.
It should be noted that collision axis goes along $z$-direcrion, impact parameter vector directed along $x$-axis and rotation
takes place around $y$ (Fig. 1).

Assuming that rotaion is the only source of the energy dependence, averaging over impact parameter and adding energy independend 
part universal for $\bar p p$ and $pp$ -- interactions,
the effect of the rotating transient matter contribution  into the directed flow, elliptic flow  and average transverse momentum 
can be calculated \cite{pt}.
It leads, in particular, 
to the following energy dependence of the average transverse momentum 
\begin{equation}\label{apte}
\langle p_T \rangle (s) = a+cs^{\delta_C}
\end{equation}
The experimental data up to the energy $\sqrt{s}=2.36$ TeV can be described well (cf. Fig. 2) using Eq. (\ref{apte}) with parameters
$a=0.337$ GeV/c, $c=6.52\cdot 10^{-3}$ GeV/c and $\delta_C=0.207$.
It should be noted that the numerical coincidence of the exponent values takes place for the average transverse momentum and 
mean multiplicity \cite{rv}. This coincidence allows one to obtain the following dependence
\[
 \langle p_T\rangle(s)=a+b\langle n\rangle (s).                                                                                              
\]
This relation is in a good agreement with experimental data at $\sqrt{s}\leq 2.36$ TeV \cite{pt}.
\begin{figure}[h]
\begin{center}
\resizebox{8cm}{!}{\includegraphics*{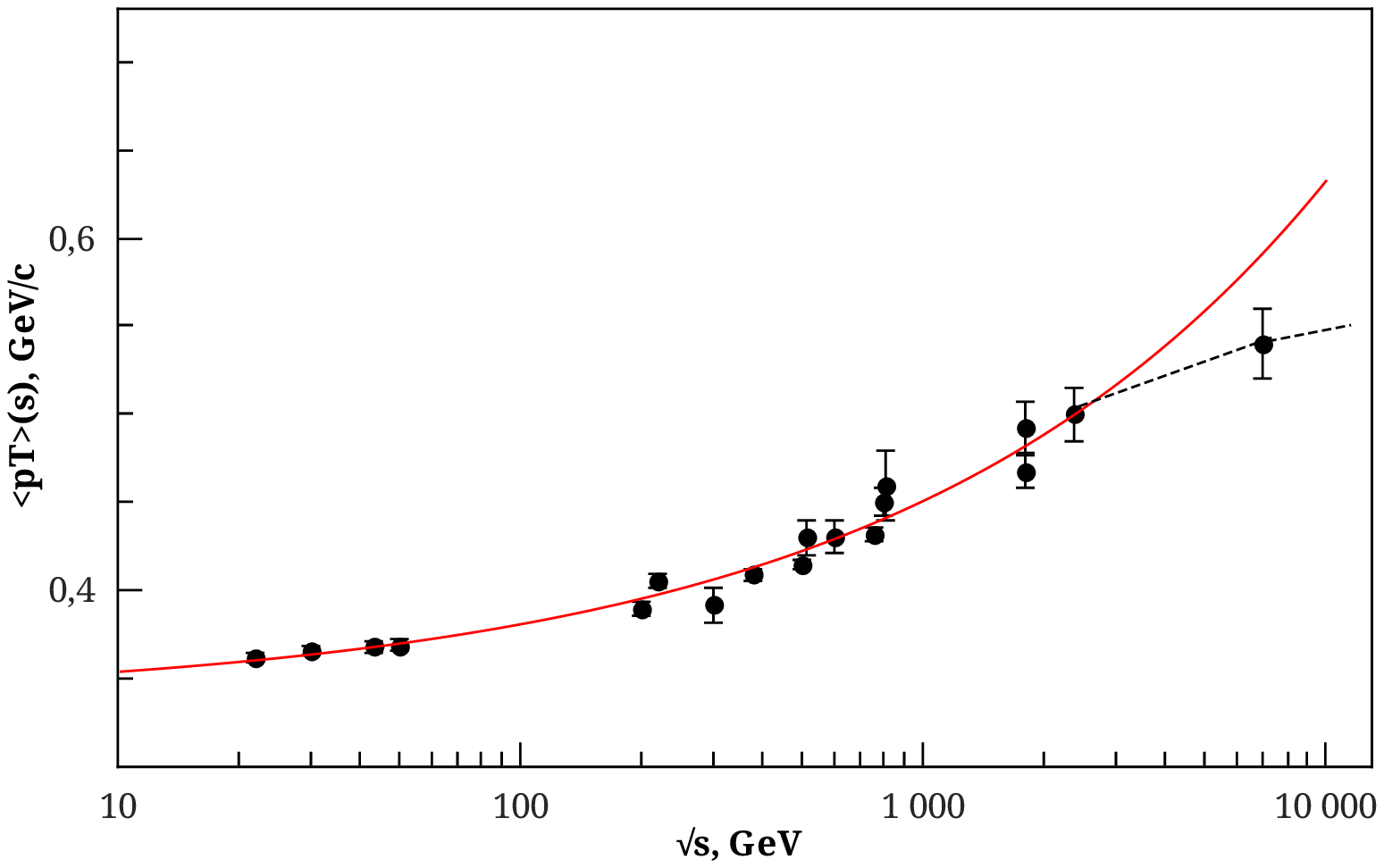}}
\end{center}
\end{figure}
Thus, this  mechanism of the average transverse momentum increase, anisotropic flows $v_1$, $v_2$ and
 two-particle correlations (ridge) generation strongly relies on the collective effect
of transient rotating matter (due to its strongly interacting nature).  
It should be noted that the dynamics of average multiplicity growth is due to the different  mechanism when a nonlinear 
field couplings   transform  the initial kinetic energy to
internal energy \cite{rv}.  
As it was noted in \cite{pt}, the formation  of the noninteracting gas of free quarks and gluons (genuine quark-gluon plasma)
would result in disappearance of the effects arising due to rotation. 
Slow down of the increase with energy and subsequent vanishing  of the energy dependent
contribution to the average transverse momentum should therefore be expected, i.e. average transverse momentum would be  reaching its
  maximum value
at some energy and then will decrease till eventual flat behavior \cite{pt}. As it was mentioned in the cited paper, no such effects were
experimentally observed by that time.

However, this situation has been changed and the recent experimental data at the highest energy $\sqrt{s}=7$ TeV \cite{cms7} started to 
demonstrate deviation from the above energy dependencies (Fig. 2).  This apparent disagreement  was mentioned in \cite{mkl} where lower values for the
exponent in the power energy dependence of the average transverse momentum have been used. It should be noted that those lower values 
for the exponent in the energy depence of the average transverse momentum result from the supposed non-universality for the $\bar pp$ and $pp$
average transverse momentum which  justification is not evident in the approch based on the color-glass condensate with universal
 transverse gluon densities  \cite{mkl}. The two different sets of the free parameters values were used in that paper.

Thus, we adhere, therefore, to the point that energy dependence of the average transverse momentum has started to change in the
 region $\sqrt{s}>3$ TeV and
this change can be interpreted as an appearance of the decoherence in the proton interactions which in its turn might results from the gradual 
phase transition of the strongly 
interacting transient state (quark-pion liquid) into the weakly interacting gas of quarks and gluons (quark-gluon plasma). Of course, further experimental
studies at higher energies are neccessary to deprive this statement of its current rather speculative flavor.

\end{document}